\documentclass{article}

\usepackage[utf8]{inputenc} 
\usepackage[T1]{fontenc}    
\usepackage{hyperref}       
\usepackage{url}            
\usepackage{booktabs}       
\usepackage{amsfonts}       
\usepackage{nicefrac}       
\usepackage{microtype}      
\usepackage{xcolor}         
\usepackage{graphicx}
\usepackage{amsmath}
\usepackage{algorithm}
\usepackage{algorithmic}
\usepackage{bbm}
\usepackage{bm}
\usepackage{subcaption}
\usepackage{wrapfig}
\usepackage[
    style=numeric,
    sorting=none,
    backend=bibtex
]{biblatex}
\usepackage{geometry}
\newcommand{\revised}[1]{#1}
\newcommand{\method}{SimAQ}

\author{%
  Jacob Egebjerg \quad Daniel Wüstner\\
  Department of Biochemistry and Molecular Biology\\
  University of Southern Denmark\\
  Campusvej 55, 5230 Odense M, Denmark \\
  \texttt{\{jegebjerg,wuestner\}@bmb.sdu.dk} \\
}

\title{SimAQ: Mitigating Experimental Artifacts in Soft X-Ray Tomography using Simulated Acquisitions}

\begin{document}

\maketitle

\begin{abstract}
Soft X-ray tomography provides detailed structural insight into whole cells but is hindered by experimental artifacts such as the missing wedge and by limited availability of annotated datasets. We present \method, a simulation pipeline that generates realistic yeast phantoms and applies synthetic imaging artifacts to produce paired noisy volumes, sinograms, and reconstructions. We validate our approach by training a neural network primarily on synthetic data and demonstrate effective few-shot and zero-shot transfer learning on real X-ray tomograms. Our model delivers accurate segmentations, enabling quantitative analysis of noisy tomograms without relying on large labeled datasets.

\end{abstract}

\section{Introduction} \label{sec:intro}

Soft X-ray tomography (SXT) is a powerful imaging modality capable of resolving the ultrastructure of intact, cryo-preserved cells in three dimensions with very high spatial resolution of 30-50 nm. SXT provides intrinsic contrast between water and carbon-rich biological material without the need for exogenous staining \cite{mcdermott2009soft,guo2019soft,carzaniga2014cryo,schneider2012cryo,loconte2023soft}. This enables quantitative imaging of cellular morphology through the linear attenuation coefficient (LAC), which reflects local molecular density and composition \cite{weinhardt2025soft}.

To recover the spatial distribution of LACs, a tomographic reconstruction is performed on tilt series of projection images acquired at successive angular increments. While full-rotation datasets can be acquired using capillary-mounted samples, many specimens of biological interest are prepared on flat grids. This geometry restricts the achievable tilt range and gives rise to the well-known "missing wedge" artifact \cite{davison1983ill}. Limited-angle acquisition introduces severe reconstruction artifacts, including elongation, and directional blurring all of which compromise the accuracy of quantitative and morphological analyses \cite{frikel2013characterization}.

In addition to geometric limitations, a range of experimental noise and artifact sources further degrade reconstruction quality. These include various sources of noise, ice cracks and voids, diffraction from impurities, non-uniform absorption due to variable ice attenuation, flat-field correction artifacts, and errors stemming from missing or misaligned projections \cite{huda2015radiographic}. Although the Filtered Back Projection (FBP) reconstruction method is commonly used due to its speed and fidelity in high signal-to-noise ratio regions, it is notably sensitive to such imperfections. Iterative reconstruction methods, such as algebraic reconstruction techniques (ART) and regularized optimization algorithms, offer improved robustness but can introduce over-smoothing and require careful hyperparameter tuning \cite{tian2011low}.

Recent advances in deep learning have introduced promising strategies for artifact correction, including supervised denoising, sinogram inpainting, and direct semantic segmentation \cite{zhao2018unsupervised, tovey2019directional, li2014dictionary, wang20200, liu2022isotropic}. These methods typically rely on supervised learning with large annotated datasets, which are time consuming to acquire and difficult to annotate, particularly when only artifact-corrupted data is available. Some approaches instead adopt self-supervised learning, for example, by rotating volumes to higher confidence angles or masking sinograms \cite{zhao2018unsupervised,tovey2019directional,liu2022isotropic}. However, the resulting training signal can be unreliable due to residual artifacts (which may be anisotropic in nature), and evaluating model performance on a complete tomogram is not feasible for data acquired on a limited-angle setup.

Here, we introduce \textit{\method}, a hybrid learning framework that leverages realistic 3D simulations of the SXT acquisition process to train a model for severe artifact correction in reconstructions and/or segmentation. We validate our approach with a neural network trained on a mix of synthetic data and sparsely annotated real SXT data. This strategy allows the model to adapt to the noise characteristics of real acquisitions while preserving the structural priors learned from simulation. In addition to improving artifact correction, the method reduces the amount of annotated data required for segmentation, even in high-quality full-range tomograms. We demonstrate the effectiveness of this approach on tomograms of \textit{Saccharomyces cerevisiae} (baker's yeast), obtaining accurate segmentations with minimal post-processing. The code for synthetic data generation and all experiments is publicly available at \url{https://github.com/Wuestner-Lab/SimAQ}.

\section{Background}  \label{sec:background}

\subsection{Principles of Soft X-ray Tomography}

Soft X-ray tomography enables high-resolution, label-free, and quantitative imaging of cryo-preserved cells and tissues without the need for staining or physical sectioning \cite{schneider2010three}. It operates within the soft X-ray water window, a spectral region between the K-absorption edges of carbon (284 eV) and oxygen (543 eV), where water is relatively transparent to X-rays while carbon-rich biomolecules absorb strongly \cite{mcdermott2009soft}. This intrinsic absorption contrast enables visualization of subcellular structures in their near-native state. Since X-ray photons penetrate much deeper than electrons, SXT can image cryo-frozen samples up to 10 \textmu m thickness, while cryo-electron tomography is limited to sample thickness of about 0.1 \textmu m. As a result, varying ice attenuation and ice-embedded structures can profoundly affect the outcome of SXT experiments. 

Image formation in SXT follows the \mbox{Beer-Lambert} law, which describes the attenuation of X-ray intensity as a function of material composition and thickness. The measured intensity $I$ after transmission through a sample is given by
\begin{equation} \label{eq:lambert-beer}
    I = I_0 \exp\left(-\int \mu(x) \, dx \right)
\end{equation}
where $I_0$ is the incident intensity and $\mu(x)$ is the spatially varying LAC. The LAC provides a quantitative, physically meaningful measure of local molecular density and composition at each voxel in the reconstructed volume, allowing for discriminating various subcellular organelles, such as lipid droplets, the vacuole or mitochondria in yeast cells \cite{weiss2000computed, uchida2011quantitative}.

Full-range SXT typically achieves isotropic resolutions on the order of 50 nm and supports the imaging of whole, intact cells under cryogenic conditions. As a result, SXT is increasingly used in structural cell biology, particularly in studies where preserving the native ultrastructure is critical \cite{loconte2023soft, weinhardt2025soft}. Typical applications include investigations of organelle organization and interaction \cite{white2020visualizing, juhl2021quantitative, loconte2022soft, egebjerg2024automated}, virus infections \cite{perez2016structural, kounatidis20203d, garriga2021imaging, loconte2021using, nahas2022near}, and morphological changes in response to environmental perturbations or genetic diseases \cite{kepsutlu2020cells, juhl2021niemann, egebjerg2024automated, szomek2024ergosterol}. The widespread use of SXT is hampered by the fact that in most set-ups, its application requires synchrotron radiation as illumination source, for which available beamtime is a limiting factor. Furthermore, the throughput is much lower than that of light microscopy, as typical acquisition times per tomogram are between 30 to 60 min, and the need for searching the grid for regions, which are suitable for imaging can easily add another hour per sample. Accordingly, the number of obtainable tomograms per imaging session is low.

\subsection{Tomographic Acquisition Constraints}
The limited availability of synchrotron beamtime and the comparably long acquisition times mean that biologically meaningful conclusions must be drawn from relatively scarce datasets. Therefore, it is crucial to minimize imaging artifacts in SXT of cells. In the following, we discuss such artifacts starting with \revised{those inherent} to the imaging process. Ideal tomographic reconstructions require projection data spanning 180$^\circ$. This can be achieved using glass capillaries as sample holders, which, however, is limited to non-adherent cells and not available at all synchrotron SXT beamlines. Furthermore, glass can absorb a significant portion of the X-rays used for imaging. Therefore, like in many other biological imaging contexts, grid-mounted samples are often used for SXT, for which the achievable tilt range is restricted due to mechanical limitations or sample geometry. This leads to incomplete coverage of \revised{the} Fourier space and a characteristic missing wedge of information \cite{davison1983ill, hanson1983bayesian}. The missing wedge exacerbates the ill-posed nature of the inverse problem and results in anisotropic resolution, systematic elongation artifacts, and reduced contrast in the affected directions \cite{davison1983ill}. 

\subsection{Experimental Artifacts}
\begin{figure}
    \centering
    \includegraphics[width=0.98\linewidth]{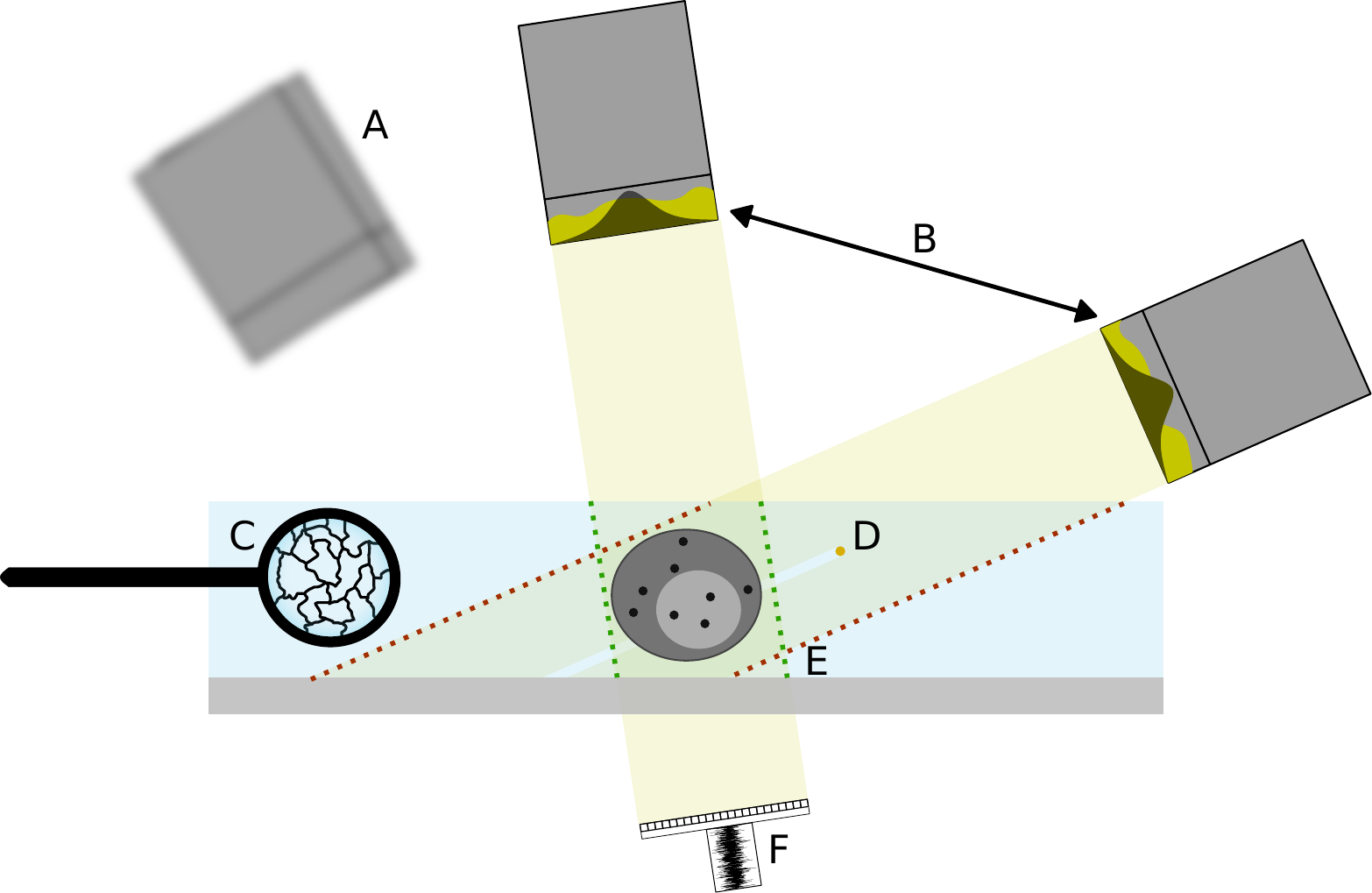}
    \caption{Examples of experimental artifact sources in cryo-SXT. A) Motion blurring resulting in missing projections. B) Flat field mismatch. The yellow flat field correction on top of the acquired projection (black), is often based on flat field images samples from elsewhere. C) Cracks in the ice. D) High absorption fiducial markers. E) Inconsistent ice attenuation (the red dashed line is \textasciitilde2.4 times further than the green in this example). F) Detector dark current. Please note that the objects are not to scale and that the X-ray source and detector are stationary, while the specimen is rotated.}
    \label{fig:artifacts}
\end{figure}

In addition to geometric limitations, SXT reconstructions are subject to a variety of experimental artifact and noise sources highlighted in Figure~\ref{fig:artifacts}, several of which are often overlooked by the tomography community. 
\begin{itemize}
    \item \textbf{Instrumental noise:} mostly arising at the detector (e.g. readout and amplification noise);
    \item \textbf{Ice damage and cracks:} which introduce structural discontinuities and scattering;
    \item \textbf{Anisotropic absorption:} especially pronounced at high tilt angles due to increased ice attenuation;
    \item \textbf{Diffraction artifacts:} caused by impurities such as ice crystals or the fiducial particles, often gold beads used for alignment;
    \item \textbf{Flat-field mismatches:} in SXT, the sample is illuminated with a focused spot, which is 'scanned' by moving the ellipsoidal mirror in a circular pattern. This generates a flickering effect and is only partially corrected for by taking a flat-field image (i.e. an image of the grid in regions without the sample) \cite{schneider2012cryo}. This correction for uneven illumination is typically carried out once but not for each projection angle, which results in systematic intensity errors across the detector.
    \item \textbf{Misalignment and missing projections:} instabilities of the sample holder can cause motion blur, and those projections have to be removed. This, together with imperfect alignment and slight tilting of the rotation axis of the sample compared to an idealized \revised{geometry, introduces} inconsistencies in the reconstruction.
\end{itemize}
These artifacts manifest non-uniformly across the volume and degrade both spatial resolution and quantitative accuracy \cite{weiss2000computed, mcnally20163d, schneider2010three}.

\subsection{Tomographic Reconstruction}
In soft X-ray tomography, the goal is to recover a three-dimensional map of $\mu(x, y, z)$ from projection images acquired at different viewing angles. Taking the negative logarithm of the transmission measurement (\ref{eq:lambert-beer}) yields the line integral of $\mu$, the \textit{Radon transform} $\mathcal{R}\{\mu\}$:
\begin{equation}
p_\theta(t) = \iint \mu(x, y) \, \delta\big(t - x \cos\theta - y \sin\theta\big) \, dx\,dy ,
\end{equation}
where $p_\theta(t)$ is the projection at view angle $\theta$ and detector coordinate $t$.

FBP approximates the inverse Radon transform through frequency-domain filtering followed by backprojection. Unlike iterative methods that introduce implicit regularization and require hyperparameter tuning, FBP preserves structures without smoothing, making it suitable when retention of fine detail is critical.

\subsection{Learning-Based Artifact Correction}

Deep learning has become a powerful tool for correcting tomographic artifacts. Approaches such as sinogram inpainting restore missing or corrupted projections \cite{zhao2018unsupervised, tovey2019directional, li2014dictionary, yao2024no}, while post-reconstruction networks map artifact-laden volumes to clean counterparts \cite{liu2022isotropic}. Encoder-decoder models, especially U-Net variants, have shown strong performance in biomedical imaging, including cryo-electron and soft X-ray tomography \cite{chen2024automated,egebjerg2024automated, erozan2024automated, dyhr20233d}. However, these methods often depend on fully supervised training with large labeled datasets, which are difficult to obtain, or on self-supervised strategies such as masking sinograms or rotating tomograms. These alternatives often fail to capture the full noise profile and often provide an unreliable training signal, especially when the noise and artifacts are anisotropic.

To mitigate this limitation, semi-supervised and domain adaptation strategies have been proposed. These methods exploit synthetic data for pretraining and use limited experimental data to refine the model's representation of real-world noise and variability. However, successfully transferring learning from synthetic to experimental domains remains a key challenge, particularly in preserving the statistical properties of noise and artifacts unique to the acquisition system.

\subsection{Related work}

Chen \textit{et al.} propose a segmentation pipeline for yeast cells using a 3D U-Net trained to segment membrane, vacuole, lipid droplet, and nucleus compartments, optimizing a cross-entropy loss function \cite{chen2024automated}. Starting from manual annotations of seven cells, they iteratively expand the training set by correcting the model’s predictions on new data. Their dataset is acquired using soft X-ray tomography (SXT) under full angular coverage with relatively low levels of noise and artifact. This also gives them access to precise LAC measurements, allowing the model to incorporate the pixel values in the decision making. Despite these favorable conditions, their model exhibits uncertainty in delineating lipid droplet boundaries, requiring post-processing refinements.

Martínez-Sánchez \textit{et al.} address the scarcity of reliable ground truth in cryo-electron tomography (cryo-ET) by introducing a simulation framework for generating biologically realistic synthetic tomograms \cite{martinez2024simulating}. Although most cryo-ET simulators focus on reproducing imaging and reconstruction artifacts, their approach prioritizes the generation of low-order cellular structures, such as membranes, macromolecular complexes, and cytoskeletal elements. The framework employs stochastic geometric models to simulate a wide variety of crowded intracellular environments. Notably, a model trained exclusively on their synthetic data was able to generalize well to real tomograms.

Yao \textit{et al.} introduce UsiNet, a deep \mbox{learning–based} method for sinogram inpainting that does not require paired ground truth data for training \cite{yao2024no}. Operating directly in sinogram space, UsiNet uses a single U-Net architecture to impute missing angular projections in an unsupervised manner, learning solely from the structure of the observed tilt series. Once completed, standard FBP is used to reconstruct a volume with significantly reduced elongation and streaking artifacts, even under large missing wedges of up to 45$^\circ$. This makes UsiNet particularly suitable for nanoparticle imaging and other materials science applications where acquisition protocols are consistent and isotropic. However, its applicability to low-contrast biological imaging remains uncertain, and the method has not been evaluated under highly anisotropic acquisition conditions typical in limited-angle SXT.

Liu \textit{et al.} propose a related self-supervised approach that operates in the reconstructed volume domain rather than sinogram space~\cite{liu2022isotropic}. Their method rotates tomograms to regions with sufficient Fourier coverage and simulates missing data by masking cones in Fourier space. This synthetic pairing allows the model to learn from known artifacts without external ground truth. They demonstrate improved reconstruction quality on several biological cryo-EM datasets, though their method also favors isotropic sampling which may limit performance in strongly anisotropic acquisition scenarios.

\section{Method} \label{sec:method}

\subsection{Simulation}\label{sec:method_sim}

\subsubsection{Phantom generation}
Yeast phantoms are simulated using ellipsoids with radii sampled from Gaussian distributions, subject to constraints that prevent organelle overlap and ensure enclosure within the cell membrane. Each phantom includes a membrane, a variable number of vacuoles, and lipid droplets. To reflect vacuole fragmentation observed in real \textit{S. cerevisiae} samples \cite{egebjerg2024automated}, some vacuoles are replaced by clusters of smaller ones generated using an Apollonian packing algorithm \cite{kasner1943apollonian}, which recursively fills gaps between mutually tangent spheres. This approach efficiently produces non-overlapping vacuole clusters with a range of sizes, without the need for explicit collision checking or costly sphere-packing simulations.

For multi-cell volumes, initial cell positions are sampled from a standard normal distribution and adjusted by minimizing a loss function that reduces overlap and promotes spatial dispersion: 
\begin{equation} \label{eq:placement_opt}
\mathcal{L} = \sum_{i=1}^{n} \sum_{j=i+1}^{n} f(r_i + r_j - \lVert \mathbf{p}_i - \mathbf{p}_j \rVert + \varepsilon) + \lambda \sum_{i=1}^{n} \lVert \mathbf{p}_i \rVert
\end{equation}
with
$$
f(x) =
\begin{cases}
x^2, & \text{if } x > 1 \\
\max(x, 0), & \text{otherwise}
\end{cases}$$
where $\mathbf{p}_i \in \mathbb{R}^2$ \revised{is} the position of cell $i$, $r_i$ its effective radius, $n$ the number of cells, and $\varepsilon$ a small constant that represents the desired gap between cells and $\lambda=0.1$. 
During optimization, cell positions are constrained to the XY plane, while Z-coordinates are sampled independently from a normal distribution, producing consistent and realistic spatial arrangements. Organelles are then voxelized at the target resolution, with intrinsic intensities drawn from Gaussian distributions based on reported LACs \cite{chen2024automated}. To reduce geometric bias from idealized shapes and introduce realistic morphological variability, random elastic deformations are applied \cite{simard2003best}.

\subsubsection{Phantom Processing}

Artifacts are introduced to the ground truth phantom through a series of controlled augmentations. Ice cracks are added using pre-generated $128\times128\times128$ volumes modulated by low-frequency Perlin noise to simulate non-uniform thickness and increase variability \cite{perlin1985image}. These volumes are precomputed to mitigate the $\mathcal{O}(N \log N)$ complexity of the generation algorithm, where $N$ is the number of voxels. The crack generation algorithm performs a seeded random walk across a 3D volume, where regions expand outward based on local weights and propagation costs. Boundaries between competing regions are extracted as crack surfaces, producing naturalistic, branching fracture patterns. Ice volumes are assembled from randomly selected precomputed samples, with the orientation of the three axes randomized independently for each sample. These are then masked to remain outside the cell membrane.

Additional high-frequency Perlin noise is added outside the membrane to simulate scattering and impurities in the surrounding ice. Fiducial markers, modeled as spheres with radii and LACs sampled from distributions matched to experimental gold beads, are inserted at random positions outside the cell membrane. The result is a noisy volume paired with a clean ground truth phantom that preserves both morphological structure and intrinsic LACs of cellular components.

\subsubsection{Projection and reconstruction}
To simulate realistic acquisition variability, the angular range is sampled from a uniform distribution (e.g. $\theta \in [70^\circ,130^\circ]$ for the experiments in Section~\ref{sec:experiments}), after which a subset of angles is randomly omitted. This models common experimental scenarios where mechanical limitations or motion artifacts render certain projections unusable. The list of retained angles is stored for use during backprojection. Gaussian additive noise is added to the angles for the forward projection to simulate error and uncertainty that may arise under experimental conditions.

Off-axis misalignment can be modeled by rolling pixel values and padding projections accordingly, though this feature was not used in our current experiments.

Flat-field correction mismatch is approximated by adding low-frequency Perlin noise, scaled by the mean intensity of each projection and a tunable hyperparameter controlling the magnitude of the flickering artifact.

The ice thickness $h$ is drawn from a Gaussian distribution, and the effective path length through ice is computed as $\ell_\text{ice}(\theta)=\frac{h}{\cos(|\theta|)}$, where $\theta$ denotes the projection angle. This increased path length at high tilt angles leads to greater attenuation, which we model as angle-dependent additive noise.

Although Poisson noise can simulate low-exposure conditions, we omit it in this work due to the high photon count of our experimental acquisitions, which renders such noise negligible.

The sinogram is reconstructed using limited-angle FBP with Ram-Lak and Hamming filters, and the output is clipped to the inscribed circle where reconstruction is theoretically valid. Limited-angle FBP introduces anisotropic distortions throughout the reconstruction due to incomplete Fourier coverage. Additionally, our real data exhibits variability in LAC distributions across acquisitions due to differences in sample preparation, ice thickness, and beam energy. To enable the network to generalize across these variations, we apply quantile normalization against a standard Gaussian distribution, which standardizes the LAC scale while preserving relative contrasts between organelles.

\begin{figure}[h]
    \centering
    \includegraphics[width=0.9\linewidth]{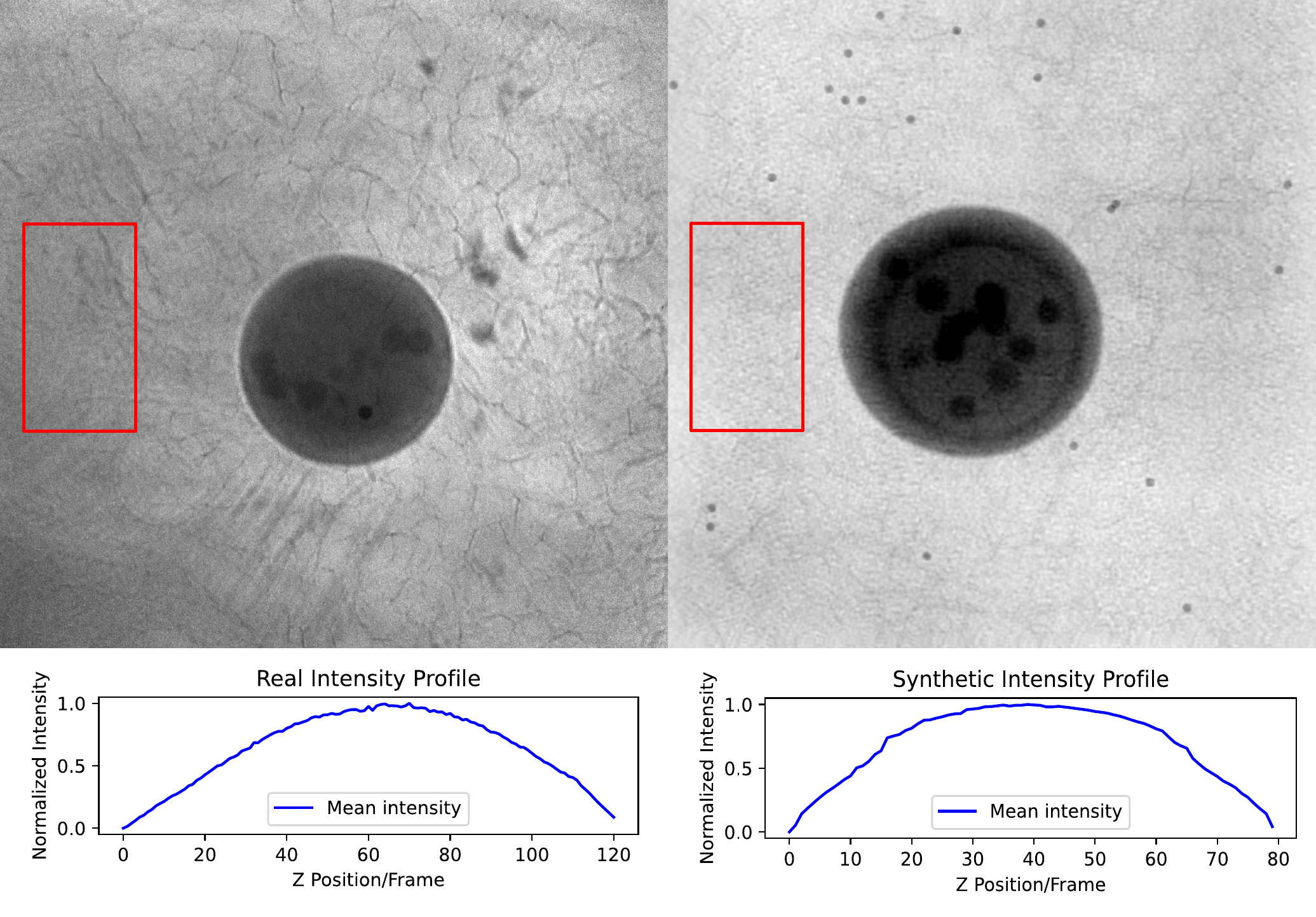}
    \caption{Projections from a real tomogram of a yeast cell acquired at \revised{BESSY} II (left) and from a synthetic tomogram with comparable structural features (right). The plotted intensity profile represents the mean pixel value within the red rectangle across all projection angles.}
    \label{fig:projections}
\end{figure}

The hyperparameters controlling the random variables in the simulation were manually tuned using known measurements from our real experimental setup and visual inspection to ensure that the resulting synthetic projections closely match real experimental data. These settings are expected to generalize well, as any additional noise beyond what is typically observed in real data mainly acts as beneficial regularization during training. We refer to the code for the full list and default values of these hyperparameters. As illustrated in Figure~\ref{fig:projections}, the simulated projection closely reproduces the structural features and noise characteristics of the real acquisition. Additional projections and corresponding sinograms are provided in Appendix~\ref{app:additional_projections}.

\subsubsection{Implementation details}

To enable efficient on-the-fly data generation for training, we prioritize computational speed without compromising realism. Although simulating the full cryo-preserved grid and surrounding ice would be more physically accurate, the associated memory requirements are prohibitive. Instead, we approximate these effects in projection space.

Voxelization is performed per ellipsoid using local bounding boxes, leveraging GPU-accelerated PyTorch tensors to evaluate the inequality $\frac{(x - x_c)^2}{a^2} + \frac{(y - y_c)^2}{b^2} + \frac{(z - z_c)^2}{c^2} \leq 1$ without wasted computation.

Forward and inverse projections are computed using the GPU-accelerated \texttt{torch-radon} library \cite{torch_radon}. To reduce the overhead of phantom generation and voxelization during training, forward and limited-angle backprojections are generated along all three principal axes, with acquisition parameters sampled independently for each orientation. During inference, only a single axis is used, as the missing-wedge artifact is specific to that projection direction.

\begin{table}[h]
\centering
\begin{tabular}{c|cccc}
\toprule
\textbf{Stage} & \textbf{1 cell} & \textbf{2 cells} & \textbf{3 cells} & \textbf{4 cells} \\
\midrule
Phantom gen    & 1.12 & 4.02 & 6.41 & 11.82 \\
Voxelization        & 0.98 & 1.81 & 2.30 & 3.21 \\
Noisy sino gen  & 10.18 & 10.38 & 10.20 & 10.08 \\
Reconstruction        & 1.11 & 1.08 & 1.07 & 1.14 \\
Normalization         & 1.00 & 0.97 & 0.97 & 0.92 \\
\bottomrule
\end{tabular}
\caption{Average runtime (seconds) per stage of the generation pipeline over 10 repetitions at $512^3$ resolution.}\label{tab:runtime}
\end{table}

On consumer-grade hardware, generating a typical volume takes approximately 15 seconds, with the exact runtime depending on the number of cells. Each run yields around 350 triplets, consisting of a phantom, its segmentation, and a noisy reconstruction. Table~\ref{tab:runtime} provides a breakdown of runtimes for different cell counts.

Phantom generation scales superlinearly with the number of cells due to the spatial organization step, which solves the optimization problem in (\ref{eq:placement_opt}). Voxelization time depends on the number of ellipsoids generated. Noise injection in phantom space is dominated by I/O overhead from loading precomputed ice crack and Perlin noise volumes, though this overhead amortizes across successive phantom generations when volumes are cached in memory. These operations scale with phantom resolution, held constant at $512 \times 512 \times 512$. The reconstruction stage, comprising forward projection, noise and artifact injection, and backprojection, scales with the number of projections. Final intensity normalization scales with volume resolution.

\subsection{Modeling}\label{sec:method_model}

\begin{figure}
    \centering
    \includegraphics[width=0.9\linewidth]{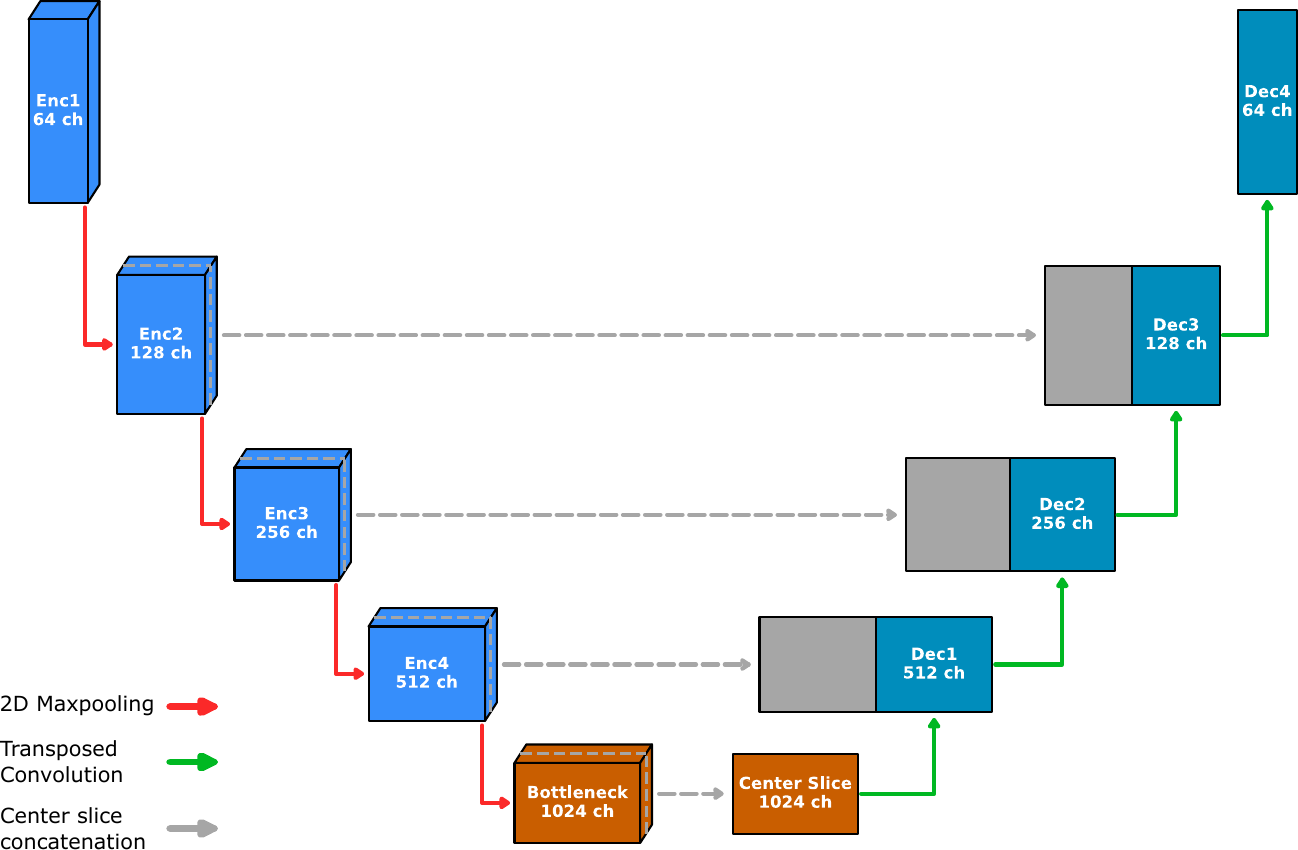}
    \caption{Hybrid 3D-to-2D U-Net architecture. The 3D encoder extracts volumetric features across multiple scales using slice-wise 2D pooling. The central slice from the bottleneck layer feeds a 2D decoder with skip connections, enabling 3D context-aware segmentation with 2D computational cost.}
    \label{fig:architecture}
\end{figure}

Using the simulation scheme described above, we generate a virtually \revised{infinite set of paired} phantoms, segmentations, and noisy reconstructions. To mitigate overfitting to the specific noise characteristics induced by simulation, we complement this synthetic dataset with a small set of partially annotated real samples (e.g. partial segmentations and no clean phantoms). Our objective is to train a neural network to recover clean phantoms or corresponding segmentations.

For the results presented in Section~\ref{sec:experiments}, we employ a hybrid 3D-to-2D U-Net architecture (see Figure~\ref{fig:architecture}) that balances volumetric context with full-field reconstruction. The model processes inputs of size $7 \times 512 \times 512$ through a four-stage 3D encoder, then decodes the central slice to produce $1 \times 512 \times 512$ outputs. The input comprises seven slices along the rotation axis, with the center slice serving as the target for prediction. This design addresses key limitations observed in alternative architectures: fully 3D U-Nets exhibited pronounced artifacts at chunk boundaries during tiled inference despite having wide receptive fields, which we attribute to limited global context in the lateral dimensions; purely 2D networks lacked sufficient depth context to consistently handle sphere edges and curved structures. We hypothesize that preserving the full lateral dimensions $(512 \times 512)$ throughout decoding is essential for mitigating missing-wedge artifacts, which manifest most prominently in the plane perpendicular to the rotation axis.

The encoder consists of four stages, each applying two 3D convolutions followed by batch normalization and \revised{PReLU (parametric rectified linear unit)} activation. Between stages, depth-preserving pooling is applied: 2D max pooling operates independently on the spatial dimensions $(W, H)$ of each slice, reducing resolution from $512 \times 512$ to $256 \times 256$ to $128 \times 128$ to $64 \times 64$ across stages, while the depth dimension $D$ remains constant at seven slices throughout the encoder. The number of feature channels doubles at each stage, starting from an initial count of 64. After the final encoder stage, the central slice is extracted from the bottleneck representation and passed to a 2D decoder.

The decoder follows a standard U-Net design with transposed convolutions for upsampling and skip connections from corresponding encoder stages. To reduce propagation of missing-wedge artifacts in the earliest reconstruction stage, the first skip connection is omitted. Each decoder block applies two 2D convolutions with batch normalization and \revised{PReLU activation}. A final $1 \times 1$ convolution produces the single-channel output. PReLU activation is used throughout to support extended training duration with high weight decay, maintaining non-zero gradients as weights are regularized \cite{he2015prelu}. The network contains approximately 68.7M parameters.

During training, the model processes seven-slice depth windows centered on the target slice, providing local volumetric context without the memory overhead of full 3D decoding. At inference, the network is applied in a sliding window fashion along the depth dimension to reconstruct entire volumes slice-by-slice. The full lateral dimensions are preserved throughout, eliminating the need for spatial tiling and avoiding chunk boundary artifacts. This architecture thus leverages volumetric context for resolving ambiguities induced by limited angular coverage while concentrating representational capacity on high-fidelity reconstruction at the native acquisition resolution.

\subsection{Transfer Learning}
Experimental sample preparation and biological variability produce highly heterogeneous data, requiring strong regularization during training to ensure generalization beyond the training distribution. We perform light fine-tuning on partially annotated samples to prevent the model from learning spurious patterns from experimental artifacts and uncharacterized debris. Following inference, predictions are post-processed with Gaussian filtering, hole filling, and volume-based filtering to remove misclassifications of extracellular debris.

Although not used in this work, several techniques could further reduce the domain gap. For example, unpaired image-to-image translation using CycleGAN \cite{zhu2017unpaired} could align real and synthetic domains without requiring additional annotations. A real-to-synthetic generator could act as a pre-processing step for segmentation, while the reverse mapping may be more appropriate when supplementing a large annotated real dataset with synthetic examples.

\section{Experiments} \label{sec:experiments}

\subsection{Dataset}
We base our experiments on a soft X-ray tomography dataset of \textit{Saccharomyces cerevisiae} (baker's yeast), acquired at the BESSY II beamline. Although the instrument nominally supports a tilt range of $\pm65^\circ$, the final 15 to 30$^\circ$ are often excluded due to alignment challenges or low signal-to-noise ratios. As a result, the effective angular coverage is typically 90 to 110$^\circ$. Projections are aligned using fiducial markers, padded with the per-projection mean, and reconstructed using FBP via Tomo3D with default parameters, including full-frequency Ram-Lak and Hamming filters. We refer to \cite{egebjerg2024automated} for more details.

We partially annotated seven tomograms from the dataset, focusing primarily on background regions where missing wedge artifacts are minimal, along with a few high-confidence pixels within cells. One volume is fully annotated across all slices, while the others contain sparse labels, emphasizing the higher-variance background. Figure~\ref{fig:annotated_data} in Appendix~\ref{app:annotated_data} illustrates the extent of these annotations with 3D renderings.

\subsection{Training}
The model described in Section~\ref{sec:method_model} is trained using a mixture of synthetic and real data, with real data comprising $15\%$ of the batch samples on average. Slices consisting only of background constitute \revised{$\sim$96\%} of all patches. To balance the training distribution, we discard these with $98\%$ probability, ensuring foreground patches dominate the training samples. For synthetic samples, the tilt range is drawn uniformly from $[70^\circ, 140^\circ]$, and the number of cells is sampled from $\max(1, \mathcal{N}(2, 1))$. The model was trained for approximately 30k steps using the Adam optimizer (learning rate $1\times10^{-4}$) with L2 regularization ($1\times10^{-5}$). A batch size of 16 was used with gradient accumulation over 4 steps (effective batch size of 64), gradient clipping at 1.0, and 16-bit Automatic Mixed Precision (AMP). Training was terminated once the loss stabilized below 0.1 for several epochs. A complete overview of hyperparameters is provided in Appendix~\ref{app:hyperparams}.

We optimize a Dice-Focal loss function, which combines the advantages of the Dice coefficient for class imbalance and the Focal loss for hard example emphasis. The loss is defined as:

\begin{align}
\mathcal{L}_{\text{Dice}} &= 1 - \frac{1}{C} \sum_{c=1}^{C}
\frac{2 \sum_i p_{i,c} g_{i,c} + \epsilon}
{\sum_i p_{i,c} + \sum_i g_{i,c} + \epsilon}, \\
\mathcal{L}_{\text{Focal}} &= - \frac{1}{N} \sum_i \sum_{c=1}^{C}
g_{i,c} (1 - p_{i,c})^{\gamma} \log(p_{i,c} + \epsilon), \\
\mathcal{L} &= \lambda_{\text{Dice}} \mathcal{L}_{\text{Dice}}
+ \lambda_{\text{Focal}} \mathcal{L}_{\text{Focal}},
\end{align}
where $p_{i,c}$ and $g_{i,c}$ denote the predicted and ground-truth probabilities for voxel $i$ and class $c$, respectively. The focusing parameter is $\gamma = 2$, and $\epsilon = 10^{-6}$ ensures numerical stability. In all experiments, we set $\lambda_{\text{Dice}} = 4$ (equal to the number of classes) and $\lambda_{\text{Focal}} = 1$, placing greater emphasis on classes with larger relative errors, typically lipid droplets, which constitute the smallest structures.

\subsection{Evaluation and Transfer to Real Data}

\begin{figure}
\centering
\includegraphics[width=1.0\linewidth]{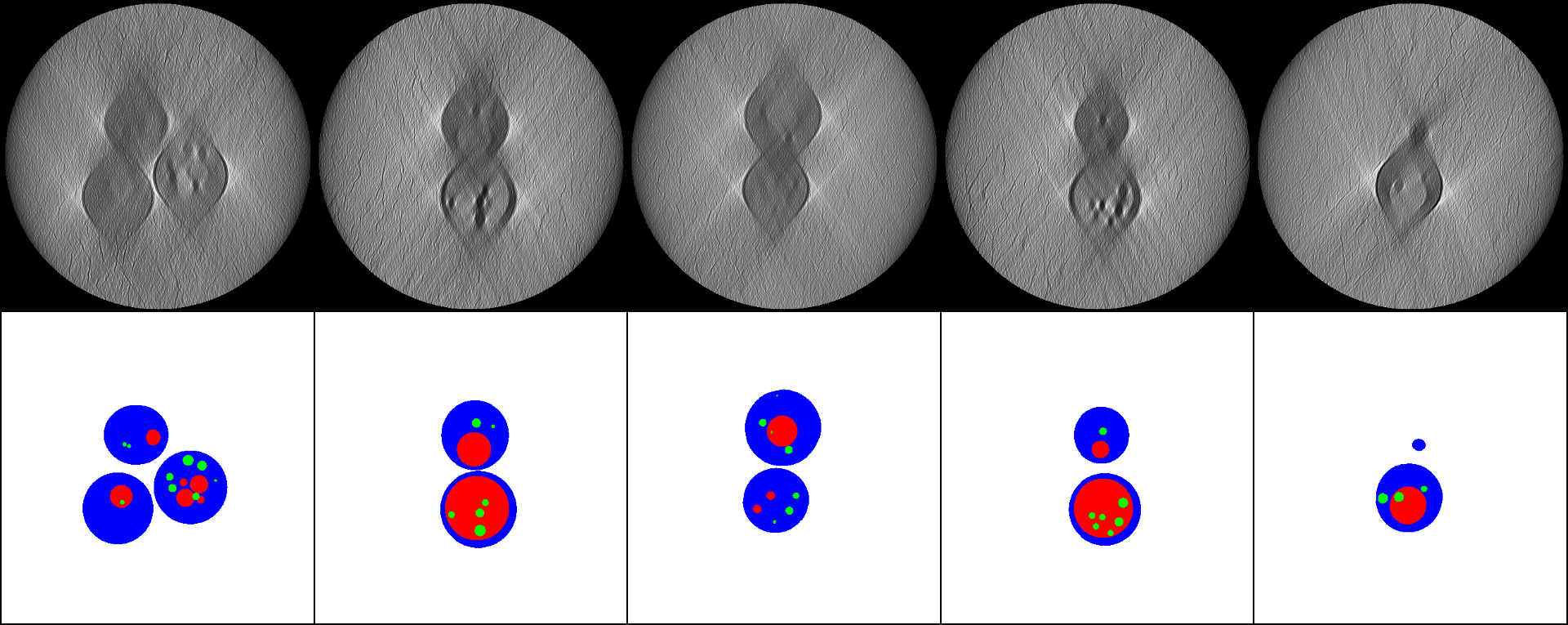}
\caption{Representative examples from the synthetic evaluation dataset. Top row: noisy input reconstructions. Bottom row: corresponding ground truth segmentations with cell membrane in blue, vacuole in red and lipid droplets in green.}
\label{fig:samples}
\end{figure}

\begin{figure}
\centering
\includegraphics[width=0.7\linewidth]{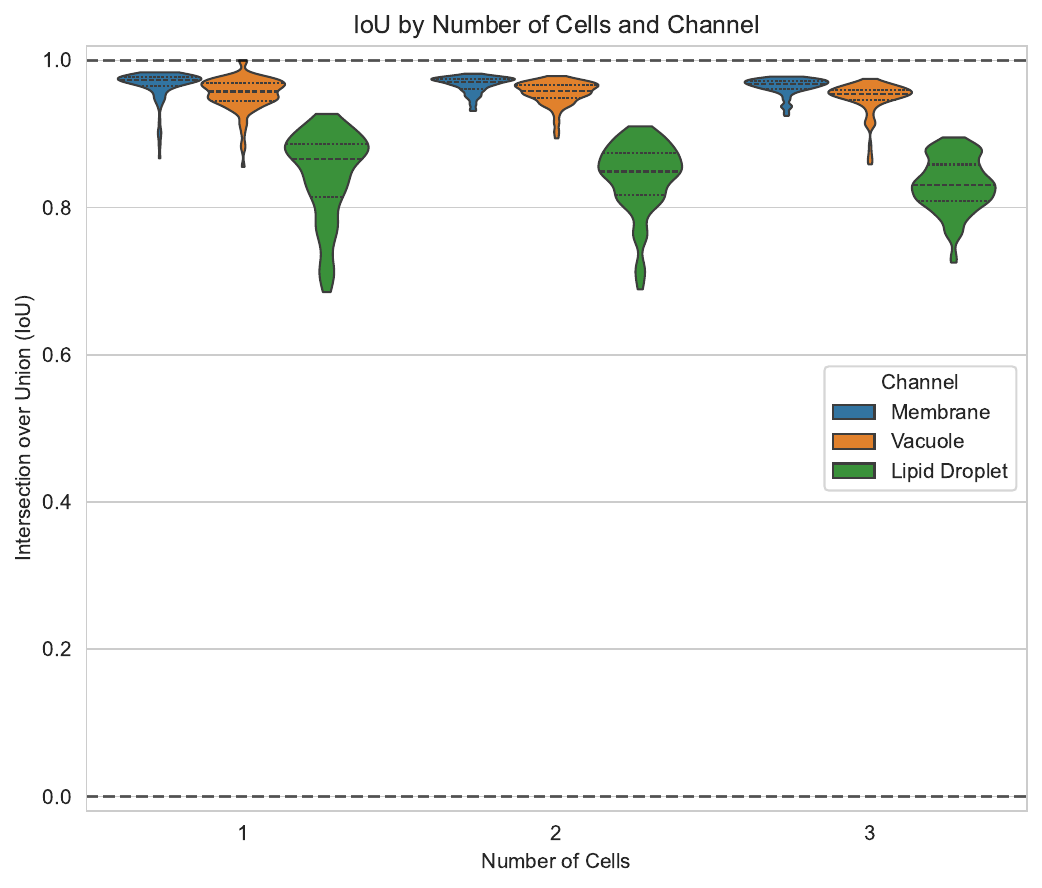}
\caption{Violin plot of IoU scores for 100 sampled tomograms across different cell counts and channels. Reconstructions are performed with an $100^\circ$ projection span. Background class has been omitted since all scores are $>0.99$\revised{.}}
\label{fig:evaluation}
\end{figure}

We quantitatively evaluate \method{} on synthetic tomograms by generating 100 volumes containing $n = 1$, $2$, and $3$ cells, with an angular span of $\theta = 100^\circ$, chosen to match the typical range (90{-}110$^\circ$) observed in our real dataset. Representative examples of the evaluation set are shown in Figure~\ref{fig:samples}, with noisy input reconstructions on the top row and corresponding ground truth segmentations on the bottom row. For each synthetic tomogram, we compute the Intersection over Union (IoU) across all four segmentation channels. The resulting distributions, shown in Figure~\ref{fig:evaluation}, demonstrate consistently high segmentation accuracy across varying cell counts. 

\begin{figure}
\centering
\includegraphics[width=0.8\linewidth]{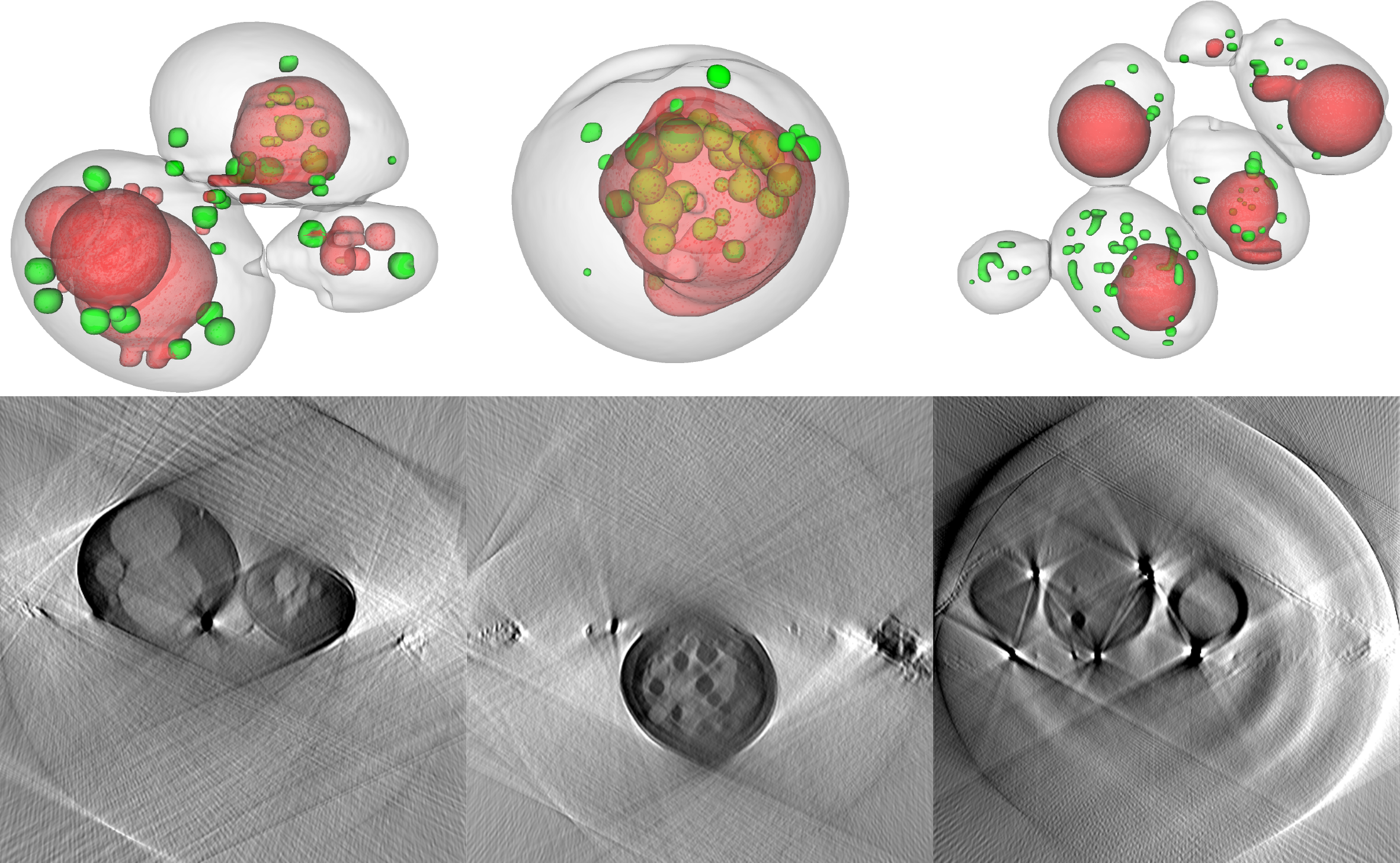}
\caption{Few-shot and zero-shot segmentation of experimental tomograms. Top: 3D renderings of model predictions. Bottom: Corresponding central slices. The model was fine-tuned using only the left two samples (left: partially annotated with labels covering regions with high confidence; middle: sparsely annotated with few annotations focusing mostly on background). The right sample was excluded from training, demonstrating zero-shot generalization. Despite minimal supervision, the model accurately segments membrane, vacuole, and lipid droplets across all cases, including the unseen tomogram.}
\label{fig:real_pred}
\end{figure}

To assess transfer learning performance, we apply the trained model to real tomograms that were either partially annotated or excluded from training. Qualitative results on five representative tomograms are shown in Figure~\ref{fig:real_pred}, illustrating realistic and accurate segmentation despite sparse annotations. These results demonstrate that \method{} can effectively generalize from synthetic data to real tomograms in both few- and zero-shot settings.

\section{Discussion} \label{sec:discussion}

A key obstacle in tomographic imaging is the lack of ground truth, which makes it difficult to assess the accuracy of reconstruction and segmentation methods. In SXT, this challenge is amplified by the inability to obtain precise annotations for cellular structures under missing wedge conditions. Our method addresses this limitation by generating realistic simulations of organelles, producing paired noisy sinograms and corresponding ground-truth volumes. This enables systematic benchmarking of algorithms under well-controlled conditions.

While we demonstrated the approach using FBP, the framework is not limited to this method. The \texttt{torch\_radon} library employed for reconstruction also supports techniques such as Landweber iteration, a common implementation of the Simultaneous Iterative Reconstruction Technique (SIRT), as well as Conjugate Gradient methods. In addition, the pipeline can operate directly in sinogram space, making it suitable for integration into sinogram inpainting workflows.

The results show that models trained primarily on synthetic data can generalize effectively to real tomograms when suitable training regularization is applied. This is achieved even in zero-shot and few-shot learning scenarios, highlighting the utility of the method for cases where annotation resources are limited. Although the experiments here focused on transfer learning, the same pipeline can support training on larger real datasets to enhance model robustness.

\subsection{Limitations}
In real data, projections are often affected by absorption and scattering from material outside the reconstruction circle or cylinder, an effect not modeled in our current synthetic pipeline. Our implementation uses yeast phantoms tailored to our experimental setup; for other setups, the phantoms must be updated accordingly. 

Additionally, the phantoms produce sharp and sometimes idealized shapes, which can differ from real cellular structures that contain additional organelles with low contrast. This simplification may limit generalization to subtle features in experimental tomograms. 

Finally, our pipeline is optimized for relatively small specimens, for which the limited depth of field of soft X-ray microscopy is negligible \cite{selin2015tomographic}. For larger specimens, simulations would need to account for out-of-focus contributions in the tomograms.

\subsection{Conclusion}
We have introduced \method, a fully three-dimensional synthetic data generation framework for tomographic reconstruction and segmentation. \method~produces realistic noisy sinograms paired with ground truth volumes and labels, enabling controlled experiments in scenarios where ground truth is otherwise unavailable. The framework is flexible, supporting multiple reconstruction algorithms and data domains. By accurately modeling cellular ultrastructure and key experimental artifacts, \method~facilitates the development of models that generalize from synthetic to real data, even in low-annotation settings. Our experiments show that models trained on \method~data achieve accurate zero- and few-shot transfer to real tomograms, enabling quantitative analysis without extensive manual labeling or complex reconstruction procedures. When adapted to a target specimen, \method~has the potential to substantially advance the development of robust and generalizable methods for SXT.

\section*{CRediT authorship contribution statement}
\textbf{Jacob Egebjerg:} Conceptualization, Data curation, Methodology, Software, Visualization, Writing - original draft. \textbf{Daniel W\"ustner:} Conceptualization, Funding acquisition, Writing - review \& editing.

\section*{Declaration of competing interest}
The authors declare that they have no known competing financial interests or personal relationships that could have appeared to influence the work reported in this paper.

\section*{Acknowledgment}
We thank Katja Thaysen and Alice Dupont Juhl for sample preparation and for sharing their yeast tomogram data. 

We are grateful to Stephan Werner and Christoph Pratsch for valuable discussions regarding the BESSY II beamline setup. We acknowledge the Helmholtz-Zentrum Berlin for providing beam time at the BESSY II facility.

Part of the computations for this project were performed on the \href{https://docs.cloud.sdu.dk/}{UCloud} interactive HPC system, managed by the eScience Center at the University of Southern Denmark.

This project is supported by the Lundbeck Foundation \mbox{(R366-2021-226)}.

\section*{Data and Code Availability}
The SimAQ simulation framework and all training code are publicly available at \url{https://github.com/Wuestner-Lab/SimAQ}. Pretrained model weights are available from the corresponding author upon reasonable request. Experimental soft X-ray tomography data are not publicly available but may be accessible through collaboration with the data holders. 

\printbibliography


\appendix
\section*{Appendix} \label{sec:app}
\renewcommand{\thesubsection}{\Alph{subsection}}

\subsection{Additional projections}\label{app:additional_projections}
\begin{figure}[h]
    \centering
    \includegraphics[width=0.8\linewidth]{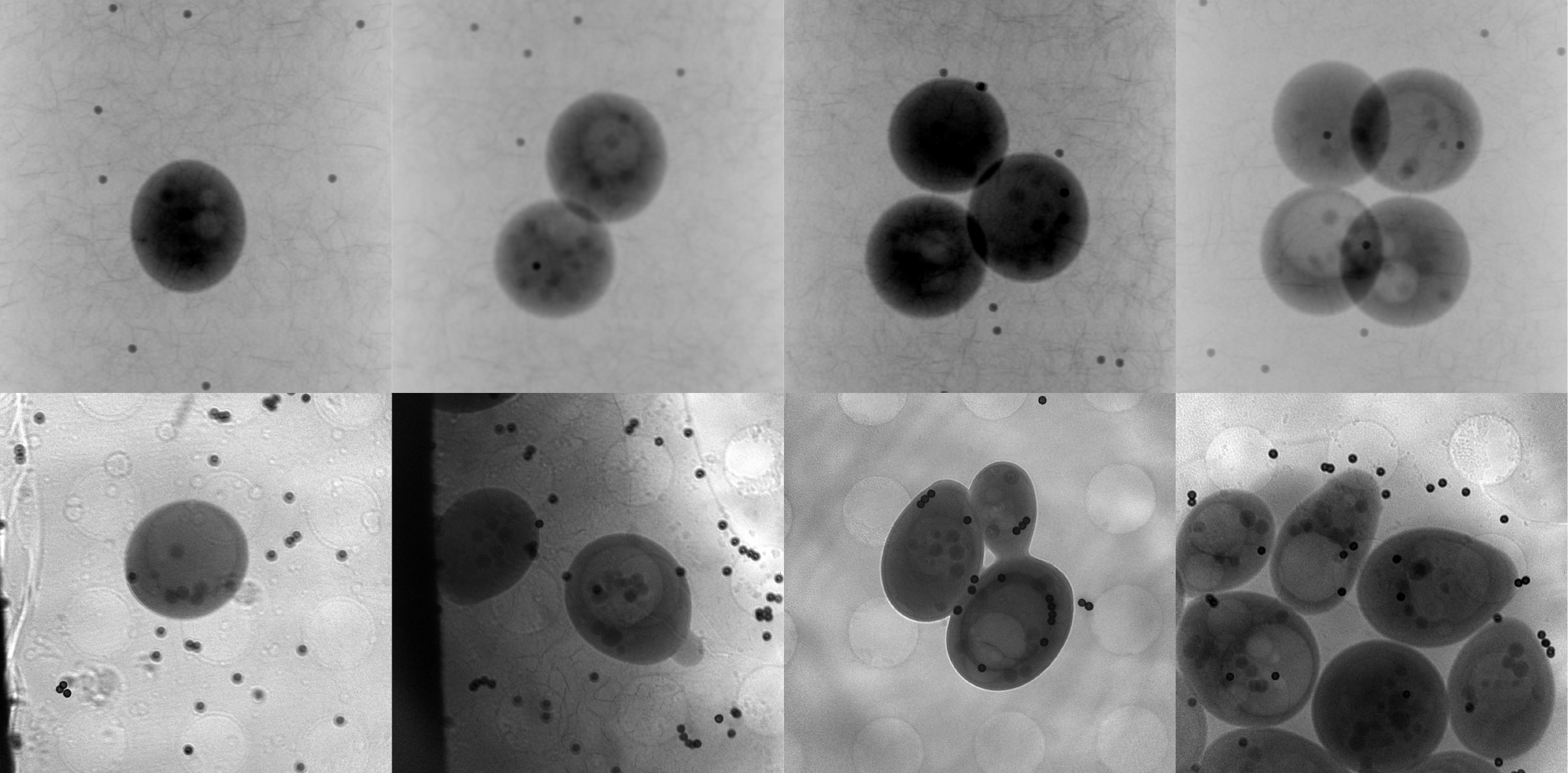}
    \caption{Comparison of projection images. Top: Synthetic projections generated from phantoms showing idealized cellular structures. Bottom: Experimental soft X-ray projections with grid (located outside reconstruction volume) and characteristic noise patterns and debris.}
    \label{fig:additional_projections}
\end{figure}

Figure~\ref{fig:additional_projections} shows representative projection images from both synthetic phantoms (top row) and experimental acquisitions (bottom row). The grid visible in real projections lies outside the reconstructed volume and does not affect the tomographic reconstruction.

\subsection{Annotated dataset}\label{app:annotated_data}
\begin{figure}[h]
    \centering
    \includegraphics[width=0.9\linewidth]{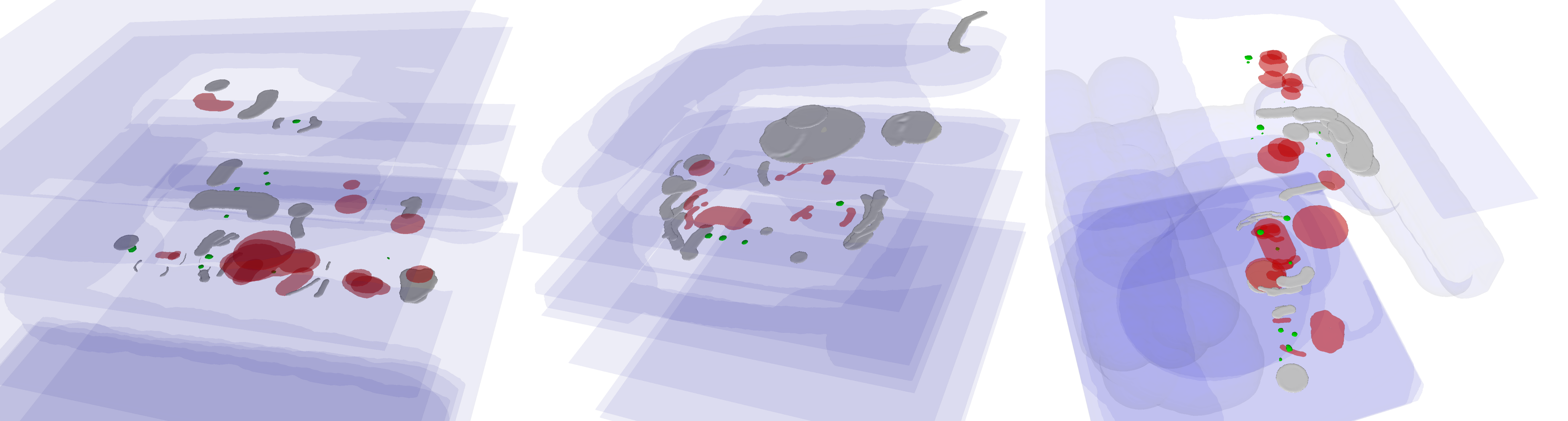}
    \caption{3D renderings of sparse manual annotations from experimental tomograms used for fine-tuning\revised{, where blue indicates background, white indicates cell membrane, red indicates vacuole, and green indicates lipid droplets}. Annotations cover small representative cellular structures.}
    \label{fig:annotated_data}
\end{figure}

Our fine-tuning dataset consists of sparse, partially annotated volumes from experimental tomograms. Figure~\ref{fig:annotated_data} shows representative 3D visualizations of the manual annotations used to adapt the model to real data characteristics and reduce overfitting to synthetic artifacts.

\subsection{Hyperparameters}\label{app:hyperparams}
Table~\ref{tab:hyperparams} lists all hyperparameters used for training the model described in Section~\ref{sec:experiments}. The real data ratio of 0.15 indicates that 15\% of each training batch consisted of real annotated data, with the remainder drawn from synthetic phantoms.

\begin{table}[h]
\centering
\begin{tabular}{ll}
\toprule
\textbf{Hyperparameter} & \textbf{Value} \\
\midrule
Conv layers per stage & 2 \\
Number of pooling layers & 4 \\
Initial features & 64 \\
Batch size & 16 \\
Gradient accumulation & 4 \\
Gradient clipping & 1.0 \\
Precision & 16-bit mixed \\
Optimizer & Adam \\
Learning rate & $1 \times 10^{-4}$ \\
Weight decay (L2) & $1 \times 10^{-5}$ \\
EMA decay & 0.999 \\
Real data ratio & 0.15 \\
Tilt range (synthetic) & $70^\circ$--$140^\circ$ (step: $1^\circ$) \\
Input depth & 7 slices \\
Input resolution & $512 \times 512$ \\
\bottomrule
\end{tabular}
\caption{Training hyperparameters for \method.}\label{tab:hyperparams}
\end{table}

\subsection{Hardware}\label{app:hardware}
Model training (Section~\ref{sec:experiments}) was performed on the UCloud interactive HPC system using a single GPU from a \texttt{u3-gpu} node. Runtime benchmarks reported in Table~\ref{tab:runtime} were measured on a consumer-grade workstation. Hardware specifications for both systems are provided in Table~\ref{tab:hardware_specs}.

\begin{table}[h]
\centering
\begin{tabular}{@{} l l @{}}
\toprule
\multicolumn{2}{@{}l}{\textbf{UCloud \texttt{u3-gpu} node (training)}} \\
\midrule
Processor & AMD EPYC 9454 @ 2.75\,GHz, 48 vCPUs \\
Memory & 192\,GB DDR5-4800 \\
GPU & NVIDIA H100 SXM5 (80\,GB) \\
\midrule
\multicolumn{2}{@{}l}{\textbf{Consumer workstation (benchmarking)}} \\
\midrule
Processor & Intel Core i9-11900 @ 2.50\,GHz (11th Gen) \\
Memory & 32\,GB DDR4-3200 \\
GPU & NVIDIA GeForce RTX 3090 (24\,GB) \\
\bottomrule
\end{tabular}
\caption{Hardware specifications for training and benchmarking.}\label{tab:hardware_specs}
\end{table}


\end{document}